\documentclass{Interspeech}
\usepackage{placeins}
\interspeechcameraready

\title{Pitfalls and Limits in Automatic Dementia Assessment}

\author[affiliation={1}]{Franziska}{Braun*}
\author[affiliation={1}]{Christopher}{Witzl*}
\author[affiliation={2}]{Andreas}{Erzigkeit}
\author[affiliation={3}]{Hartmut}{Lehfeld}
\author[affiliation={3}]{Thomas}{Hillemacher}
\author[affiliation={1}]{Tobias}{Bocklet}
\author[affiliation={1}]{Korbinian}{Riedhammer}

\affiliation{}{Technische Hochschule Nürnberg}{Germany}
\affiliation{}{Geromed GmbH Erlangen}{Germany}
\affiliation{Klinik für Psychiatrie und Psychotherapie}{Universitätsklinik der Paracelsus Medizinischen Privatuniversität, Klinikum Nürnberg}{Germany}
\email{franziska.braun@th-nuernberg.de, witzlch88229@th-nuernberg.de}

\usepackage{comment}
\usepackage{pgfplots}
\usepackage{pgfplotstable}
\usepackage{xcolor}
\usepackage{tikz}

\begin{document}

\maketitle

% the abstract here must exactly match the abstract entered into the paper submission system
\begin{abstract}
Current work on speech-based dementia assessment focuses on either feature extraction to predict assessment scales, or on the automation of existing test procedures. Most research uses public data unquestioningly and rarely performs a detailed error analysis, focusing primarily on numerical performance. We perform an in-depth analysis of an automated standardized dementia assessment, the Syndrom-Kurz-Test. We find that while there is a high overall correlation with human annotators, due to certain artifacts, we observe high correlations for the severely impaired individuals, which is less true for the healthy or mildly impaired ones. Speech production decreases with cognitive decline, leading to overoptimistic correlations when test scoring relies on word naming. Depending on the test design, fallback handling introduces further biases that favor certain groups. These pitfalls remain independent of group distributions in datasets and require differentiated analysis of target groups.

%    - andere: auf basis von cookie-theft 
%    - ptifalls: corr hoch != funktioniert gut (für alle gruppen), vor allem Memory
%    - limits: 
%        - wenn nichts gesagt wird, erkennt der Erkenner nichts (darf sowas ins Endergebnis mit einfließen?), sind die Endkorrelation nicht zu optimistisch aufgrund dessen 
%        - skt8: alle vorlesen... dadurch treffer limit der automatischen auswertung ("unkooperativer user"
%        - skt9: Nutzer "torpediert" Testausführung, d.h. dass der Nutzer den Anweisungen nicht folgt (nur die Symbole nennt, an die er sich erinnern kann) und daher die Auswertung nicht einfach möglich ist
\end{abstract}

\renewcommand{\thefootnote}{\fnsymbol{footnote}}
\footnotetext[1]{Equal contribution.}

\section{Introduction}
%The demographic shift in most countries has led to a notable rise in age-related cognitive disorders such as Alzheimer’s disease and other forms of dementia. 
%Consequently, there is an increasing need to detect these conditions as early as possible. 
In clinical practice, dementia screening usually comprises a combination of medical and psychological assessments, including history taking, biomarker extraction, and standardized cognitive testing. 
These assessments are typically administered by clinical experts after self or medical indication (e.g., decline in cognitive abilities).
However, neuropsychological testing is time-intensive as it needs professionally supported assessment of a broad range of cognitive capabilities such as short- and long-term memory, recognition, and numerical understanding. 
This increases the costs of diagnosis and limits the feasibility of early detection initiatives.
As a result, providing prompt identification of these conditions, and enabling dementia care accessibility, remains hardly given for many persons affected.
Recent advances in automated speech recognition (ASR), make automatic cognitive assessments based on speech possible, holding the potential to facilitate early diagnosis, enable medical implementation, and enhance the quality of life for those affected. 
Moreover, data collection in automatic settings would help with studying larger populations for the early stages of cognitive decline, improving both research and clinical outcomes.
Since many established neuropsychological tests rely heavily on verbal test responses, they appear suitable for ASR-based automation. 
However, state-of-the-art ASR systems such as OpenAI's Whisper \cite{radford2022robustspeechrecognitionlargescale} exhibit multiple challenges when deployed in cognitive testing scenarios. 
The target speech domain comprises elderly speech (which differs from the middle-aged speech on which ASR systems are mostly trained), increased dialectal variations and accents within older populations and among dementia patients, as well as pathological speech patterns such as prolonged pauses and interruptions. 
Further challenges arise from the structured and constrained nature of speech elicited during cognitive testing, which differs significantly from spontaneous, natural speech, which ASR systems are usually trained on (e.g., podcast and video sources).
For effective early dementia detection and monitoring, automated test evaluation must remain highly sensitive to subgroups with no or mild cognitive impairment, while screening for later dementia stages remains relatively straightforward. 
Tackling previously named challenges, this work analyses the speech-based automation of the Syndrom-Kurz-Test (SKT), an established cognitive performance test that is sensitive to the early stages of dementia syndrome.
We automatically evaluate the SKT subtest based on manual and automatic transcripts using OpenAI's open-weight ASR model Whisper.
In a case study, we analyze the overall correlation to expert scoring for 158 subjects performing the SKT as part of a comprehensive dementia assessment procedure.
Further analysis of correlations for individual subgroups of individuals with no cognitive impairment (NCI), mild cognitive impairment (MCI), and dementia (DEM) reveals potential pitfalls in automatic speech-based dementia assessment.
We find that evaluation performance decreases for NCI and MCI groups and is related to high transcription word error rates.
In an in-depth test analysis, we discuss the pitfalls and limits of automatic dementia assessment, which can lead to over-optimistic results and misleading conclusions. 
We also propose guidelines to mitigate these issues, providing the basis for more interpretable, transparent, and reliable automated dementia assessment.

\section{Related Work}

Speech-based dementia assessment has garnered increasing attention in recent years, with research efforts mainly focusing on two key approaches. 
The first approach involves extracting linguistic and acoustic features from elicited speech (e.g., picture description tasks) to estimate cognitive impairment through established clinical scales such as the Mini-Mental State Examination (MMSE) and the Montreal Cognitive Assessment (MoCA) \cite{adress20,adresso21,madress22,taukadial24,braun_classifying_2023}. 
The second one centers on automating neuropsychological test procedures (e.g., Verbal Fluency, Boston Naming Test, CERAD-NB) by leveraging speech processing technologies \cite{troeger18,koenig18,kim19,kwon21,lofgren_breaking_2022,braun22_interspeech,braun_GoingCookieTheft_2022}. 
Although these methods have shown promise, they often rely on limited publicly available datasets. The DementiaBank Pitt corpus \cite{becker1994dementiabank}, consisting of U.S. English speech data, remains the most widely used resource. 
Initiatives such as the ADReSS \cite{adress20}, ADReSSo \cite{adresso21}, and MADReSS \cite{madress22} challenges have demonstrated that it is possible to differentiate between pathological and non-pathological speech samples.
However, the clinical relevance of this binary classification is limited, as even laypersons may distinguish dementia patients from healthy controls.
Commonly employed dementia screening tools, such as the MMSE \cite{mmse75} and MoCA \cite{moca_2005}, are widely used due to their simplicity and ease of administration. 
Nonetheless, these tools face several limitations, including a lack of normalization for demographic factors like age and education, and a binary classification of cognitive status.
In contrast, the Syndrom-Kurz-Test \cite{erzigkeit77, erzigkeit15} offers a more detailed assessment by classifying cognitive performance across six stages based on attention and memory tasks, making it more sensitive to early-stage dementia.
In our previous work, we demonstrated the feasibility of automating SKT scoring on a small subset of 30 subjects using \cite{braun22_interspeech}, revealing strong correlations with expert assessments when using manual transcripts.
The reliability of automated evaluations decreased using automatic transcripts due to recognition errors of the ASR system (mod9), particularly in SKT subtests 2, 6, and 9.
Expanding upon previous research, our work emphasizes a subgroup-specific analysis of automatic dementia assessments, focusing on differences in ASR accuracy across individuals with no cognitive impairment (NCI), mild cognitive impairment (MCI), and dementia (DEM). 
By considering these subgroups, we aim to uncover potential biases and limitations inherent in automatic speech-based dementia screening methods.

\section{SKT}
\label{sec:SKT}
The Syndrom-Kurz-Test (SKT) \cite{erzigkeit77}, is a test instrument for assessing memory and attention deficits,
%, which is available in various languages (e.g., German, English, French, Spanish, and many more). 
%The SKT is 
characterized by its patient-centered design and brief administration time, %making 
it is particularly well-suited for routine clinical practice.
Furthermore, the test’s individual tasks are designed for clarity and can be explained using alternative phrasing or different languages, ensuring broad accessibility.
The test inventory comprises nine subtests, derived from established psychological and psychiatric assessment methods. 
The subtests are categorized based on their primary focus: either attention or memory performance. 
For providing a comprehensive understanding of the test’s execution and evaluation, please refer to our work \cite{braun22_interspeech, erzigkeit15}.
Test performance in the attention subtests (1, 3, 6, 7) is assessed based on the time required to complete the respective tasks, while the memory subtests (2, 8, 9) are evaluated by counting the number of omitted objects; these concepts are used to derive the raw test scores. 
In attention subtests, subjects are instructed to complete the tasks as quickly and accurately as possible. 
Errors need to be pointed out by the examiner and require the subjects' correction; the correction time is included in the processing time. 
% In memory subtests, participants are given one minute to recall the memorized objects, as speed is not a primary evaluation criterion. 
In the memory sub-tests, it is not necessary to work as quickly as possible; the time available should be sufficient to recall the objects.
% eindampfen?
However, if an incorrect object is named, the examiner asks whether the participant is certain about their response. 
%If the incorrect response is confirmed, it will be recorded as a confabulation.
%While confabulations and other response details are recorded in the test protocol, they do not directly contribute to the final test score.
Confirmed misnamings are logged as confabulations, which are noted in the protocol but not scored.
The normalization of raw test scores accounts for subjects' age and IQ range (\textless90, 90–110, \textgreater110), resulting in norm scores between 0 and 3 for each subtest. 
The sum of all norm values yields the SKT total score, ranging from 0 to 27. 
This total score serves as an indicator of cognitive impairment, particularly when memory and attention sub-scores exhibit homogeneous profiles \cite{erzigkeit77}.
The interpretation of the overall SKT result follows a clinically established classification. 
Total scores between 0 and 4 indicate no cognitive impairment, while scores between 5 and 8 correspond to mild cognitive impairment. 
For dementia severity, the total score differentiates between mild (9--13), moderate (14--18), severe (19--23) and very severe dementia (24--27).
Regardless of the overall score, the sub-scores for memory and attention provide a detailed assessment of impairment in the respective cognitive domain.

\section{Data}
This study builds on a subset of the Nuremberg single-center corpus \cite{braun22_interspeech}, comprising audio recordings along with ground truth transcriptions following the guidelines of Dresing \& Pehl \cite{dresing_pehl10}. 
These recordings capture clinical interviews, as well as SKT and CERAD-NB assessments performed at the Memory Clinic of the Department of Psychiatry and Psychotherapy at Nuremberg Hospital.
The recordings were conducted as part of routine clinical practice, where all subgroups (NCI, MCI, and DEM) were recruited from the same pool of patients mitigating selection and interviewer bias. 
It should be noted that some participants exhibit strong regional dialects and accents, and all speakers wore surgical masks, both holding challenges for ASR systems.
The subset consists of 158 individuals aged between 49 and 89 years ($\mu = 73.69 \pm 9.02$), including 63 male and 95 female participants doing the SKT. 
The cohort presents a diverse range of pre-existing conditions, such as depression, dementia, and other neurological or psychiatric disorders. 
Table \ref{tab:cohortCharacteristics} summarizes the demographics of the subgroups NCI, MCI, and DEM based on the SKT total score. 
%``NCI'' denotes either no cognitive impairment or only age-related memory decline, while ``DEM'' encompasses both mild and moderate cases of the condition.
All tests were conducted and evaluated in face-to-face sessions by psychological experts, strictly following the SKT Manual guidelines \cite{erzigkeit77} (see section \ref{sec:SKT}). 
During the evaluation process, raw scores from the memory and attention subtests were normalized using the established SKT norms from 2015 (see Table 11 in \cite{erzigkeit15}).

\begin{table}[htbp]
  \centering
  \caption{Demographics for NCI, MCI, and DEM subgroups.}
  \label{tab:cohortCharacteristics}
  \begin{tabular}{l||c|c|c|c}
    \toprule
    \textbf{Group} & \textbf{N} & \textbf{Age} & \textbf{Gender} & \textbf{Education Years} \\
    \midrule
    \textbf{NCI}      & 68  & 73.8$\pm$8.9  & 34m/34f& 12.9$\pm$2.2 \\
    \textbf{MCI}     & 44  & 74.2$\pm$10.1 & 16m/28f& 12.3$\pm$2.8 \\
    \textbf{DEM}     & 46  & 73.1$\pm$8.2  & 13m/33f& 11.5$\pm$2.0 \\
    \midrule
    \textbf{Overall} & 158 & 73.7$\pm$9.0 & 63m/95f& 12.3$\pm$2.4 \\
    \bottomrule
  \end{tabular}
\end{table}

\section{Method}

\subsection{Automatic Transcription}
    The ASR system ``Whisper'' (\texttt{whisper-large-v3}) is used for automatic transcription; model weights are open-source and can be accessed online\footnote{\url{https://huggingface.co/openai/whisper-large-v3}}.
    Inference is performed with the open-source runtime environment \texttt{whisper.cpp}\footnote{\url{https://github.com/ggerganov/whisper.cpp/releases/tag/v1.7.4}}, with the language set to German and timestamp generation disabled.
    Voice Activity Detection (VAD) is used for pre-processing the raw audio to isolate speech segments and remove non-speech parts such as silence, background noise, and other irrelevant artifacts. 
    This is particularly important for encoder-decoder based ASR systems, which may otherwise produce hallucinated output in silent or low-energy segments. 
    To improve segmentation accuracy, a smoothing strategy is applied that includes a padding mechanism to maintain linguistic coherence and prevent abrupt segment transitions.
    Following VAD-based segmentation, different chunking strategies are explored to optimize transcription accuracy and coherence. 
    The first approach involves merging all detected speech segments into a single continuous audio file before transcription (\textbf{merged}). 
    This ensures that context is maintained throughout the recording though it could introduce artifacts when merging disjointed speech segments. 
    The second approach processes each detected speech segment independently, storing and transcribing them separately (\textbf{chunked}). 
    While this method allows for precise alignment between transcriptions and their corresponding timestamps, it risks losing global context, particularly for long-form transcription.
    To mitigate the loss of contextual information, a third approach combines independent chunking with a prompting mechanism. 
    Here, transcriptions from previous chunks are used as prompts for subsequent chunks, helping to maintain coherence across segment boundaries, particularly in extended speech interactions (\textbf{prompted}). 

\subsection{Automatic Scoring}

    The analysis focuses exclusively on subtests that can be evaluated based on an audio recording and its corresponding transcript. 
    To ensure accurate number recognition, all numbers in the transcripts are normalized to their digit equivalents (e.g., ``eins'' to ``1'', for numbers between 0 and 99). 
    This process accounts for dialectal variations (e.g., ``zwei'' vs. ``zwo'' mapped to ``2'').
    Furthermore, to obtain accurate word timestamps for scoring, forced alignment is performed using the Kaldi-based \texttt{mod9 engine}\footnote{\url{https://mod9.io}}, which utilizes DNNs and HMMs for transcription and alignment.
    Test evaluation is performed separately for each task group. 
    The first group includes subtests 1 (object naming), 2 (immediate recall of objects), 8 (delayed recall of objects), and 9 (object recognition). 
    In these tasks, the number of correctly recognized objects -including synonyms- is determined using a Levenshtein distance-based similarity measure (90\% match threshold relative to word length). 
    The recognized words are then stored along with their respective timestamps. 
    For subtest 1, the processing duration is determined using the timestamp of the last recognized word, which serves as the score. 
    For subtests 2, 8, and 9, the number of missing objects is computed based on the recognized items.
    The second group comprises subtests 3 (reading numbers), 6 (counting symbols), and 7 (interference test). 
    Here, the extracted words from the transcript are compared against the expected task-specific responses, including predefined two-digit numbers, the expected range of counted symbols, and the intended letters. 
    The timestamp of the last correctly identified word within the expected response set is used to determine the processing duration, which serves as the final score. 
    If no valid number or interference letter is detected in the transcript, it is assumed that the participant did not respond, and the raw score was set to 60.

\section{Experiments}
Transcription quality is measured by Word Error Rate (WER) and Word Correctness (WC) with respect to human ground truth transcriptions.
Automated test scoring is evaluated using Pearson correlation with the expert-assigned scores.
Metrics were calculated by selecting the best audio pre-processing for each subtest and for each subject group. 
This is intended to investigate performance across different patient groups in automatic evaluation and transcription, as well as to allow differentiated interpretation.

\subsection{Target Group Analysis} 
\tablename~\ref{tab:skt_best_audio_correlation} shows the Person correlation for the automatic evaluation of the SKT based on ground truth (GT) and automatic (ASR) transcriptions.
We obtain high ASR correlations for subtests 1, 2, 3, 8, and 9 approaching or even exceeding GT correlations (e.g., for SKT1), suggesting that the automatic evaluation of these subtests is straightforward.
It is noticeable that 8 performs better than 2 despite sharing the same task (recalling objects) and that 6 and 7 show lower performance for ASR as well as GT, leading us to deeper analysis.
%At first sight, this would suggest that the fully automatic evaluation is working well, and in some cases even better than the evaluation based on the manually produced transcriptions.

\begin{table}[htbp]
  \centering
  %\caption{Correlation (Pearson) for evaluation of SKT subtests based on ground-truth (GT) and ASR-based transcription, along with audio pre-processing method. \emph{Overall} refers to a single correlation computed across all pooled data points. Best correlation per subtest is underlined for readability.}
  \caption{Pearson correlation with expert scores for SKT subtest evaluation using ground-truth (GT) and ASR transcripts with different audio pre-processing. \emph{Overall} is computed across all pooled data points.}
  \label{tab:skt_best_audio_correlation}
  \begin{tabular}{c||c|c|c}
    \toprule
    \textbf{Subtest} & \textbf{GT Corr.} & \textbf{ASR Corr.} & \textbf{Preprocessing Type} \\
    \midrule
    1       & 0.82             & \underline{0.93}  & prompted \\
    2       & \underline{0.86} & 0.84              & merged   \\
    3       & 0.85 & \underline{0.88}             & merged   \\
    6       & \underline{0.79} & 0.47              & merged   \\
    7       & \underline{0.87} & 0.74              & chunked  \\
    8       & \underline{0.97} & 0.95              & merged   \\
    9       & 0.72             & \underline{0.73}  & raw      \\
    \midrule
    \textbf{Overall} & \underline{0.86}       & 0.77  & ---      \\
    \bottomrule
  \end{tabular}
\end{table}
\FloatBarrier

%Next, the individual target groups are examined in more detail, with the group of NCI and MCI patients being the most important; dementia patients are also important, but the focus here lies more on monitoring. 
%For this purpose, people are assigned to one of the 3 groups based on the expert assessment described in section \ref{sec:SKT}. 
%We observe that the performance between the groups is quite different, especially the evaluation quality for unimpaired people is quite low in comparison, see  \tablename~\ref{tab:cohortCharacteristics}.
%The overall correlation of 0.76 can therefore be deceptive, especially if critical groups, as here, only have a comparatively low correlation score.

The individual target groups are analyzed, focusing primarily on NCI and MCI subjects, while DEM subjects are considered mainly for monitoring. 
Based on expert assessments (Section \ref{sec:SKT}), participants are assigned to one of three groups. 
Performance varies notably between groups, with unimpaired individuals showing comparatively low evaluation quality (\tablename~\ref{tab:correlations_groups}). The overall correlation of 0.77 (\tablename~\ref{tab:skt_best_audio_correlation}) may therefore be overoptimistic while critical groups exhibit lower performance.

\begin{table}[htbp]
  \centering
  %\caption{Pearson correlation values for each subtest across cognitive groups. \emph{Overall} refers to a single correlation computed across all pooled data points. Best correlation per subtest and group is underlined for readability.}
  \caption{Pearson correlation with expert scores per subtest and cognitive group. \emph{Overall} is computed across pooled data points.}
  \label{tab:correlations_groups}
  \begin{tabular}{c||c|c|c|c|c|c}
    \toprule
    \textbf{Subtest} & \multicolumn{2}{c}{\textbf{NCI (n=68)}} & \multicolumn{2}{c}{\textbf{MCI (n=44)}} & \multicolumn{2}{c}{\textbf{DEM (n=46)}} \\
    \cmidrule(r){2-3} \cmidrule(r){4-5} \cmidrule(r){6-7}
                     & GT & ASR  & GT & ASR  & GT & ASR \\
    \midrule
    1 & 0.72 & 1.00 & 0.62 & 0.97 & 0.96 & 0.88 \\
    2 & 0.69 & 0.78 & 0.85 & 0.78 & 0.90 & 0.73 \\
    3 & 0.72 & 0.75 & 0.73 & 0.73 & 0.96 & 0.99 \\
    6 & 0.61 & 0.28 & 0.82 & 0.41 & 0.70 & 0.65 \\
    7 & 0.58 & 0.18 & 0.80 & 0.76 & 0.81 & 0.75 \\
    8 & 0.87 & 0.90 & 1.00 & 0.94 & 0.97 & 0.91 \\
    9 & 0.49 & 0.42 & 0.61 & 0.77 & 0.66 & 0.68 \\
    \midrule
    \textbf{Overall} & 0.70 & 0.51 & 0.82 & 0.78 & 0.87 & 0.82 \\
    \bottomrule
  \end{tabular}
\end{table}
\FloatBarrier

Analyzing individual tests, such as subtest 7, reveals pitfalls due to multiple factors.
The NCI group shows a high WER and low WC, indicating frequent transcription errors unlike in the MCI and DEM groups. 
This stems from more impaired individuals often using the full 60 seconds with pausing, aiding speech recognition, while less impaired individuals (NCI) speak faster and more fluently, leading to repetition hallucinations in the specific atypical speech context (i.e., sequence of letters), resulting in lower correlation and metrics.

\begin{table}[htbp]
  \centering
  \caption{
  %Pearson Correlation, WER, and WC values for subtest 7 across groups. GT depicts evaluation using manually created transcripts while ASR shows the results using generated transcripts.
  Pearson correlation, WER, and WC for subtest 7 across groups; GT uses manual transcripts, ASR uses generated ones. \emph{All} refers to a single correlation computed across all data.
  }
  \label{tab:correlation_results_skt7}
  \begin{tabular}{l||c|c|c|c}
    \toprule
    \textbf{Group}  & \textbf{GT} & \textbf{ASR} & \textbf{WER} & \textbf{WC} \\
    \midrule
    NCI       & 0.58 & 0.18 & 1.05 & 0.49 \\
    MCI      & 0.80 & 0.76 & 0.53 & 0.50 \\
    DEM & 0.81 & 0.75 & 0.63 & 0.55 \\
    \midrule
    All      & 0.87 & 0.74 & 0.78 & 0.51 \\
    \bottomrule
  \end{tabular}
\end{table}
\FloatBarrier

\tablename~\ref{tab:correlation_results_skt2_skt8} shows correlation differences between subtests 2 and 8, both requiring object recall. 
Although subtest 8 shows a higher correlation, this is not reflected in the WER and WC values.
This stems from the SKT structure  and the behavior of the subjects between the two tests;  
subtest 2 follows the immediate stimulus (subtest 1) directly where subjects tend to memorize more items (immediate recall), making speech recognition more impactful, whereas delayed recall in subtest 8 reduces this effect since fewer items are memorized. 

\begin{table}[htbp]
  \centering
  %\caption{Pearson Correlation, WER, and WC values for subtests 2 and 8 across groups. GT depicts evaluation using manually created transcripts while ASR shows the results using generated transcripts. \emph{All} refers to a single correlation computed across all pooled data points.}
  \caption{Pearson correlation, WER, and WC for subtests 2 and 8 by group. ASR uses generated transcripts. \emph{All} refers to a single correlation computed across all pooled data points.}

  \label{tab:correlation_results_skt2_skt8}
  \begin{tabular}{l||c|c|c|c|c|c}
    \toprule
    \textbf{Group} & \multicolumn{3}{c|}{\textbf{SKT2}} & \multicolumn{3}{c}{\textbf{SKT8}} \\
    \cmidrule(r){2-4} \cmidrule(r){5-7}
                   & \textbf{ASR} & \textbf{WER} & \textbf{WC} & \textbf{ASR} & \textbf{WER} & \textbf{WC} \\
    \midrule
    NCI  & 0.78 & 0.35 & 0.68 & 0.90 & 0.34 & 0.70 \\
    MCI & 0.78 & 0.35 & 0.66 & 0.94 & 0.40 & 0.62 \\
    DEM & 0.73 & 0.46 & 0.63 & 0.91 & 0.49 & 0.53 \\
    \midrule
    All & 0.81 & 0.38 & 0.66 & 0.92 & 0.40 & 0.63 \\
    \bottomrule
  \end{tabular}
\end{table}
\FloatBarrier

Subtest 9 (see \tablename~\ref{tab:correlation_results_skt9}) shows correlation differences within the subgroups; NCI is scoring lowest (0.42), followed by DEM (0.68) and MCI (0.77), despite lower performance in form of WERs and WCs with increasing impairment. 
Healthier individuals tend to recall more items, making speech recognition performance more critical. 
Further data analysis revealed three subjects who only pointed instead of naming objects, and others who verbally confirmed or denied objects, introducing errors in automatic evaluation.

\begin{table}[htbp]
  \centering
  \caption{
  %Pearson Correlation, WER, and WC values for subtest 9 across groups. GT depicts evaluation using manual created transcripts while ASR shows the results using generated transcripts.
  Pearson correlation, WER, and WC for subtest 9 by groups; GT uses manual transcripts, ASR uses generated ones. \emph{All} refers to a single correlation computed across all pooled data points.
  }
  \label{tab:correlation_results_skt9}
  \begin{tabular}{l||c|c|c|c}
    \toprule
    \textbf{Group} & \textbf{GT} & \textbf{ASR} & \textbf{WER} & \textbf{WC} \\
    \midrule
    NCI  & 0.49 & 0.42 & 0.30 & 0.74 \\
    MCI & 0.61 & 0.77 & 0.31 & 0.72 \\
    DEM & 0.66 & 0.68 & 0.38 & 0.67 \\
    \midrule
    All & 0.72 & 0.73 & 0.33 & 0.71 \\
    \bottomrule
  \end{tabular}
\end{table}
% \FloatBarrier

\subsection{Adressing Limitations}

To mitigate the previous limitations and improve subtest-specific performance, we have taken the following adaptations:
For subtest 6 scoring, we use the last word timestamp, for subtest 7, we employ a grammar-supported mod9 engine for ASR, and for subtest 9 use the LLM \texttt{Qwen2.5-32B-Instruct-MLX-4bit}\footnote{\url{https://huggingface.co/lmstudio-community/Qwen2.5-32B-Instruct-MLX-4bit}} to filter negated words. 
These methods raise the overall correlation from 0.77 to 0.87 (compare \tablename~\ref{tab:skt_best_audio_correlation} and \tablename~\ref{tab:correlations_overall}). Subtest 6 correlations increase notably for MCI (0.41 to 0.74) and DEM (0.65 to 0.74), while subtest 9 shows limited improvement in the DEM group due to LLM unreliability. Subtest 7 correlations improve significantly (NCI: 0.18 to 0.91) with mod9, though results rely on therapist-intervened letter production. Overall, NCI and MCI group correlations increase from 0.51 to 0.68 and 0.78 to 0.87, respectively, emphasizing the importance of precise subgroup analysis and adaptations.

\begin{table}[htbp]
  \centering
  \caption{Pearson Correlation values for different SKT subtests across cognitive subgroups and overall.}
  \label{tab:correlations_overall}
  \begin{tabular}{c||c|c|c|c}
    \toprule
    \textbf{Subtest} & \textbf{NC} & \textbf{MCI} & \textbf{DEM} & \textbf{Overall} \\
    \midrule
    1 & 1.00 & 0.97 & 0.88 & 0.93 \\
    2 & 0.78 & 0.78 & 0.73 & 0.84 \\
    3 & 0.75 & 0.73 & 0.99 & 0.88 \\
    6 & 0.29 & 0.74 & 0.88 & 0.74 \\
    7 & 0.91 & 0.95 & 0.88 & 0.96 \\
    8 & 0.90 & 0.94 & 0.91 & 0.95 \\
    9 & 0.41 & 0.77 & 0.68 & 0.73 \\
    \midrule
    \textbf{Overall} & 0.68 & 0.87 & 0.88 & 0.87 \\
    \bottomrule
  \end{tabular}
\end{table}
\FloatBarrier

% \subsection{ADReSS Challenge Evaluation}

% %\tablename~\ref{tab:word_count_dementia_classification} demonstrates once again that it is essential to look closely at the data/test environment. 
% %In this extreme example, classifications are made based only on the number of words spoken to determine whether someone has dementia or not. 
% %Although, interestingly, this classification can be done more precisely if the number of words spoken by the interviewer is included, taking into account the age of the patient - even achieving up to 73\% accuracy.

% \tablename~\ref{tab:word_count_dementia_classification} highlights the importance of careful data and test environment analysis. 
% In this extreme case, dementia classification is based solely on word count, achieving up to 73\% accuracy when including interviewer word count and subject's age. 
% However, this approach lacks clinical validity, as high accuracy from such simplistic features does not reflect meaningful cognitive assessment.

% \begin{table}[htbp]
%   \centering
%   \caption{Classification accuracy using investigator (INV) and participant (PAR) feature sets.}
%   \label{tab:word_count_dementia_classification}
%   \begin{tabular}{lcc}
%     \toprule
%     \textbf{Speaker} & \textbf{Features}      & \textbf{Accuracy} \\
%     \midrule
%     INV & word\_count           & 0.68 \\
%     INV & word\_count, age      & \underline{0.73} \\
%     PAR & word\_count           & 0.64 \\
%     PAR & word\_count, age      & 0.55 \\
%     \bottomrule
%   \end{tabular}
% \end{table}
% \FloatBarrier

\section{Conclusion}
This study underscores the critical importance of detailed target group and error analysis in the automatic assessment of dementia and other medical conditions. 
Without such analyses, there is a risk of drawing overoptimistic conclusions about the reliability of automated methods, obscuring group-specific weaknesses that could compromise clinical effectiveness.
Supplementary experiments of predicting dementia in the ADReSS Challenge using only word count and age of the interviewer or patient, resulted in high accuracies of up to 0.73, highlighting the need for the interpretability and transparency of concepts underlying the numerical performance.
This is particularly concerning in clinical and assistive contexts, where accurate transcription is essential for diagnosis, communication aids, and research.
To enhance the robustness of automatic speech recognition in these domains, future research should consider hybrid approaches that integrate Whisper's capabilities with traditional ASR methods. Such strategies may improve performance consistency across diverse patient populations. Furthermore, we strongly advocate for explainable and differentiated evaluation methodologies, as demonstrated by our findings, which reveal that superficial performance improvements do not always translate to meaningful diagnostic advancements. Ensuring methodological transparency and comprehensive analysis will be crucial in maintaining the validity and comparability of future research in this field.

\pagebreak

\bibliographystyle{IEEEtran}
\bibliography{mybib}

\end{document}